\documentclass[runningheads,a4paper]{llncs}

\usepackage{amssymb}
\usepackage{graphicx}

\begin{document}

\mainmatter  
\title{Flexible protein-peptide docking using CABS-dock with knowledge about the binding site }
\titlerunning{Flexible protein-peptide docking using CABS-dock}
\author{Mateusz Kurcinski, Maciej Pawe\l{} Ciemny, Maciej Blaszczyk, Andrzej Kolinski and Sebastian Kmiecik}
\authorrunning{Flexible protein-peptide docking using CABS-dock}
\institute{University of Warsaw, Chemistry Department,\\ Pasteura 1, 02-093 Warsaw,\\ Poland
}
\maketitle
\begin{abstract}
Despite considerable efforts, structural prediction of protein-peptide complexes is still a very challenging task, mainly due to two reasons: high flexibility of the peptides and transient character of their interactions with proteins. Recently we have developed an automated web server CABS-dock (http://biocomp.chem.uw.edu.pl/CABSdock), which conducts flexible protein-peptide docking without any knowledge about the binding site. Our method allows for full flexibility of the peptide, whereas the flexibility of the receptor is restricted to near native conformations considering the main chain, and full flexibility of the side chains. Performance of the CABS-dock server was thoroughly tested on a benchmark of 171 test cases, both bound and unbound. Evaluation of the obtained results showed overall good performance of the method, especially that no information of the binding site was used. From unsuccessful experiments we learned that the accuracy of docking might be significantly improved, if only little information of the binding site was considered. In fact, in real-life applications user typically has access to some data indicating the location and/or structure of the binding site. In the current work, we test and demonstrate the performance of the CABS-dock server with two new features. The first one allows to utilize the knowledge about receptor residue(s) constituting the binding site, and the second one allows to enforce the desired secondary structure on the peptide structure. Based on the given example, we observe significant improvement of the docking accuracy in comparison to the default CABS-dock mode. 
\end{abstract}

\section{Introduction}

Peptides are probably one of the best candidates for protein-targeting drugs as they are highly selective and effective on one hand and easily tolerated and rather safe on the other. Recently, the interest in research and development of the peptide-based drugs has been constantly on the rise [1, 2]. Consequently, there is a large need for effective tools and strategies for computer-aided peptide modeling that may help to establish new routes of rational peptide design [3].

For over ten years we have been developing a coarse-grained model of proteins CABS [4] and applying it to various molecular modeling tasks including modeling of protein structure [5, 6], dynamics [7-12] and interactions [13-20]. Recently, we have launched the CABS-dock web server [21, 22], which utilizes a method for protein-peptide docking through multiscale simulations using CABS model. What distinguishes the CABS-dock from other docking protocols is the fact that no information about the shape and location of the binding site on the surface of the protein receptor is used. Method was tested on a large benchmark of protein-peptide complexes, both bound and unbound [21, 22]. The results showed overall good performance of the CABS-dock. What we also learned from the less successful cases is that the docking results might significantly improve if some information about the binding site and/or the approximate shape of the bounded peptide was used. Both these sources of data reduce dramatically the conformational space of the system. This way the search for the global energetic minimum is far more effective.

Here we present initial validation tests of new features that could be added to the CABS-dock method. These features utilize additional data about the modeled complex, such as approximate location of the binding site or secondary structure of the peptide. We also demonstrate how this data improve the modeling of the complex between the Syntenin protein tandem and a peptide (pdb code: 1W9E).

\section{Materials and Methods}
\subsection{Model}

The CABS model has been already described in great detail [4]. Here we only outline its main characteristics: 1) coarse-grained representation of proteins and peptides -- each residue is represented by three pseudo-atoms: \textbf{C}arbon \textbf{A}lpha, carbon \textbf{B}eta and united atom for the \textbf{S}ide-chain (hence the name of the model). Additional fourth center of interactions is located in the center of the virtual C$\alpha$-C$\alpha$ bond and mimics the hydrogen bonding point of the peptide bond. To speed-up the computations of local conformational transitions C$\alpha$ atoms may only move between nods of the simple cubic lattice, while other atoms follow the movement of the main chain; 2) statistical force field -- interactions contribute to the total energy of the system in relation to how often they have been observed in already solved structures from the PDB database; 3) simulation is controlled by the Replica Exchange Monte Carlo scheme. Thanks to such design the CABS simulations are almost four orders of magnitude faster than all-atom Molecular Dynamics. At the same time CABS Monte Carlo dynamics simulations preserve acceptable resolution of the modeled structures, as coarse-grained models may be easily and accurately rebuilt to an all-atom representation.

\subsection{CABS-dock Standard Procedure}

Peptide docking procedure implemented in the CABS-dock server consists of the following steps: 1) initial setup -- both the peptide and the receptor are transformed into CABS representation, ten copies of the peptide are generated for the Replica Exchange method, each copy has a random structure and is located in a random spot around the receptor structure at a distance up to 20\AA~from the receptor surface; 2) simulation -- both the main chain and the side groups of the receptor are flexible however the whole structure is restricted to near native conformations (although user may select semi-flexible and fully-flexible spots), the peptide molecules are freely moving around the receptor and are completely flexible; 3) initial filtering -- as a result of the simulation 10000 poses are collected, top 1000 most strongly bound complexes are selected for the next step; 4) clustering -- 1000 models are grouped into 10 clusters in k-medoids procedure with ligand RMSD (root mean square deviation of coordinates of the peptide after superposition of the receptor molecules) as the measure of models similarity, medoids from each cluster are selected for the next step; 5) model reconstruction and final ranking -- reduced models are rebuild to the all-atom representation using Modeller [23] and ranked 1-10 based on the cluster density (maximal difference between models within a cluster divided by the number of cluster elements) of the cluster from which they originate.

\subsection{New Features of the CABS-dock Method}

We have designed and implemented two additional CABS-dock features, which utilize additional knowledge about the modeled complex.

\subsubsection{Docking with anchoring residues.}

This new CABS-dock feature allows user to select receptor residue(s) that potentially belong to the binding site (preferably on the surface of the protein) and will act as anchor(s) upon docking. That functionality was realized by a simple attractive potential between anchoring residues and any of a peptide residues. The energy of such interactions depends only on the minimum distance $D$ (Fig. 1) between any of the peptide’s side-chains and the side-chain of the anchoring residue. The potential drops to zero both when $D> D_{max}$ to avoid attraction through the core of the receptor, but also when $D<D_{min}$ so that the shape of the energy function close to the binding site is not distorted. Such interaction has a broader meaning than a Go-like contact potential. This concept follows a hypothesis that different residues of the receptor may be responsible for peptide recognition and for binding [24].

\begin{figure}
\centering
\includegraphics[height=6.2cm]{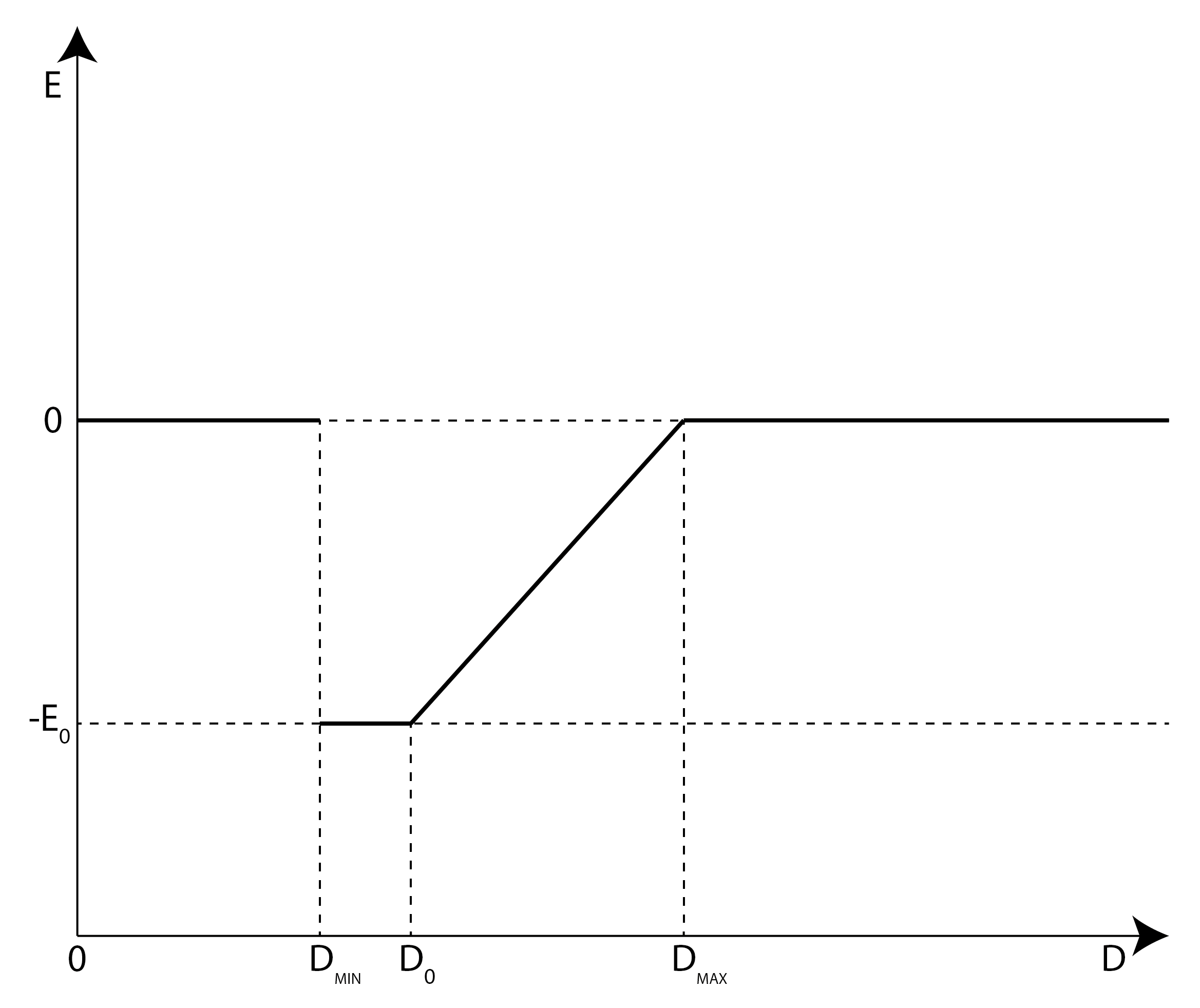}
\caption{Plot of the attractive potential between anchoring residues and the peptide.  The potential is equal to $-E_0$ on the range defined by $D_{MIN}$ and $D_0$. Then it is linearly driven to zero at the distance $D_{MAX}$. }
\label{fig1}
\end{figure}

Using anchoring residues may significantly improve the quality of docking. If user provides the location of the binding site there is no need to sample other regions around the receptor. Therefore, we also updated the procedure for building of the initial setup of the system. Instead of “randomly anywhere on the sphere” around the receptor, the peptide copies are initially located on the same sphere but “randomly in the vicinity” of the intersection of that sphere and its radius defined by the geometrical center of all anchoring residues. 

\subsubsection{Docking with enforced secondary structure}
 
This CABS-dock feature allows user to enforce predicted or known secondary structure of the peptide molecule. In the standard mode, the CABS-dock algorithm uses secondary structure prediction only as a weak preference towards helical or expanded conformation -- there is a small energetic reward when peptide adopts structure compatible with the prediction, but no penalty if it does not. We added a procedure that generates distance constraints for those fragments of the peptide that are assigned helix or strand. The lengths of the constraints are taken from average helix/strand structure.

\section{Results}

For demonstrational purposes, we present here one case in which the use of additional information about the peptide made a dramatic difference in the results (thorough testing of the new features will be performed on a large benchmark set and will include the optimization of various parameters).

Protein complex 1w9e consists of the two-domain receptor (165 residues) and a 5-residue peptide (sequence: NEFYF). In three CABS-dock prediction runs (3x10000 models) the best model generated had ligand RMSD vs. the native structure equal to 3.77\AA, but the best pose from those selected as the final models was 16.41\AA~away from the native structure [22]. Therefore, in the benchmark test this case was considered as a low-quality prediction.

We conducted three simulations using new CABS-dock features: 1) only with enforced secondary structure, 2) only with anchoring residues 3) with both enforced secondary structure and anchoring residues. The secondary structure was assigned to the peptide via DSSP [25] as CCEEC meaning that residues 2-4 in the peptide were constrained (two residues flanking the beta fragment, although assigned coil are affected by neighboring strand and therefore also constrained). As the anchoring residues we selected those amino acids that were found to form a contact with any of the ligand residues. We considered two residues to be in contact if any of their heavy atoms were located within 4.5\AA~from each other. For the comparison with the benchmark result we selected the best models from 10 final models from each of the simulations.

\begin{figure}
\centering
\includegraphics[height=6.2cm]{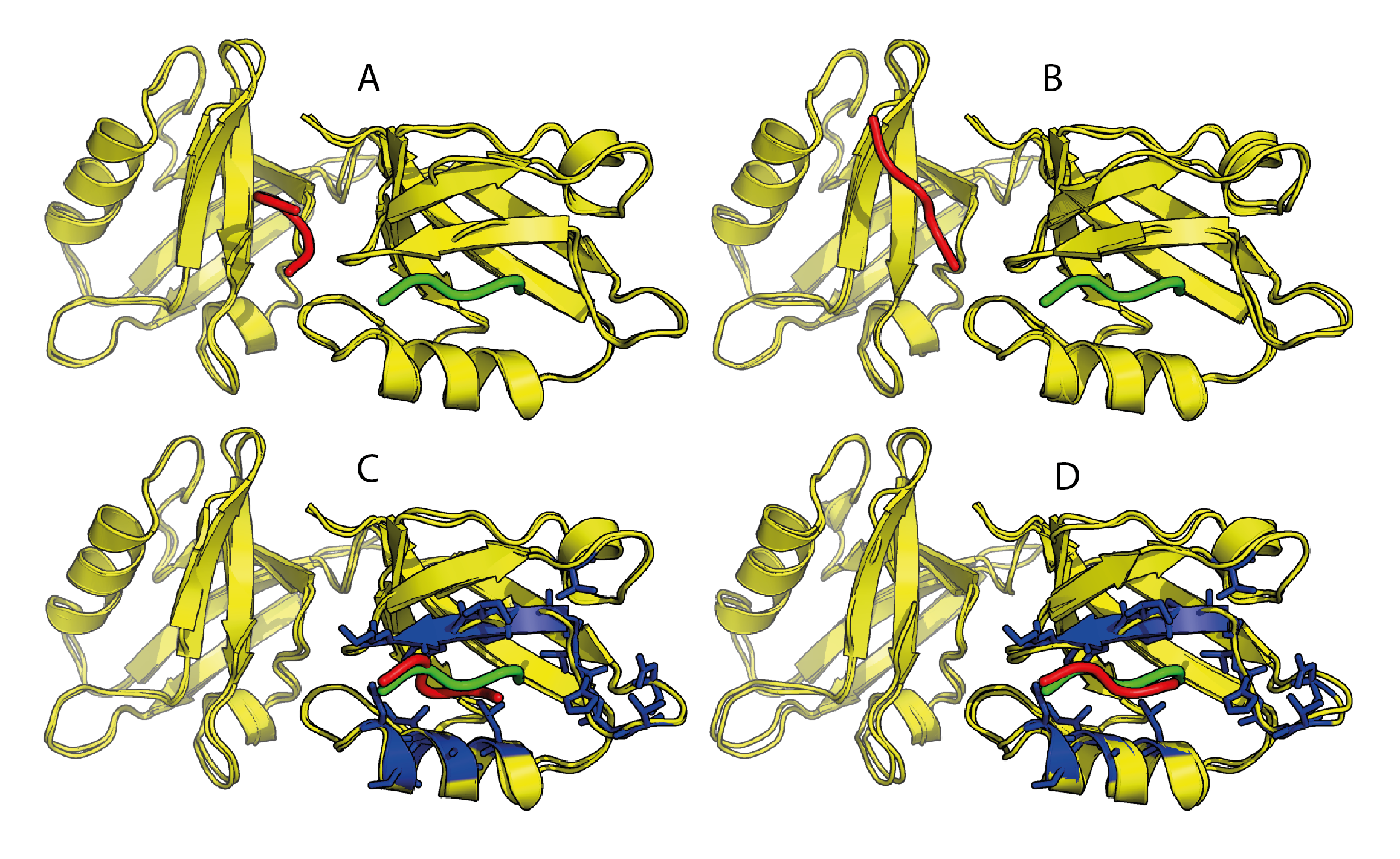}
\caption{Docking results of a) standard CABS-dock simulation (resulting RMSD of the structure is 16.41\AA)  b) with enforced secondary structure (RMSD of the structure is 21.25\AA) c) with anchoring residues (RMSD 3.51\AA) d) with enforced secondary structure and anchoring residues (RMSD  0.85\AA). Models are superimposed on the experimental complex structure. The receptor is shown in yellow, peptide is shown in green (native) and red (models). Anchoring residues are shown in blue.}
\label{fig2}
\end{figure}

\subsubsection{Simulation 1.} Use of only the enforced secondary structure feature resulted in the RMSD rise to 21.25\AA. Similarly to the benchmark simulation the peptide was docked in the wrong binding site. However as a result of the enforcement of the secondary structure, the peptide was more linear and therefore more similar to the native conformation (RMSD between peptides after their superposition was equal to 1.19\AA), see Fig. 2b.

\subsubsection{Simulation 2.} In this case the anchoring residues pulled the peptide to the correct pocket as reflected in the RMSD drop to 3.51\AA. The shape of the peptide molecule however differs visibly from the native conformation (RMSD between peptides after their superposition -- 2.34\AA), see Fig. 2c.

\subsubsection{Simulation 3.} With both features combined the resulting model is only 0.85~\AA \\away from the native structure and the peptide structure is 0.43\AA~away from the peptide in the native structure, see Fig 2d.

\section{Conclusions}

New features designed for the CABS-dock server greatly improved docking results of the test case complex 1w9e. Obviously further optimization of these procedures followed by broader testing on a larger sample is required to draw conclusions about the effectiveness of the changes. Moreover both features seem to be very sensitive to false positives. Nevertheless the spectacular level of improvement in the test case is a solid base for further development. 

\section{Acknowledgments}

The authors acknowledge support from the National Science Center grant [MAESTRO 2014/14/A/ST6/00088]


\begin{thebibliography}{99}
\bibitem{} Tsomaia, N., Peptide therapeutics: targeting the undruggable space. Eur J Med Chem, 2015. 94: p. 459-70.
\bibitem{} Fosgerau, K. and T. Hoffmann, Peptide therapeutics: current status and future directions. Drug Discov Today, 2015. 20(1): p. 122-8.
\bibitem{} Audie, J. and J. Swanson, Recent work in the development and application of protein-peptide docking. Future Med Chem, 2012. 4(12): p. 1619-44.
\bibitem{} Kolinski, A., Protein modeling and structure prediction with a reduced representation. Acta Biochim Pol, 2004. 51(2): p. 349-71.
\bibitem{} Blaszczyk, M., et al., CABS-fold: Server for the de novo and consensus-based prediction of protein structure. Nucleic Acids Res, 2013. 41(Web Server issue): p. W406-11.
\bibitem{} Kmiecik, S., M. Jamroz, and M. Kolinski, Structure prediction of the second extracellular loop in G-protein-coupled receptors. Biophys J, 2014. 106(11): p. 2408-16.
\bibitem{} Jamroz, M., A. Kolinski, and S. Kmiecik, CABS-flex predictions of protein flexibility compared with NMR ensembles. Bioinformatics, 2014. 30(15): p. 2150-4.
\bibitem{} Jamroz, M., A. Kolinski, and S. Kmiecik, Protocols for efficient simulations of long-time protein dynamics using coarse-grained CABS model. Methods Mol Biol, 2014. 1137: p. 235-50.
\bibitem{} Jamroz, M., A. Kolinski, and S. Kmiecik, CABS-flex: Server for fast simulation of protein structure fluctuations. Nucleic Acids Res, 2013. 41(Web Server issue): p. W427-31.
\bibitem{} Kmiecik, S. and A. Kolinski, Characterization of protein-folding pathways by reduced-space modeling. Proc Natl Acad Sci U S A, 2007. 104(30): p. 12330-5.
\bibitem{} Wabik, J., et al., Combining coarse-grained protein models with replica-exchange all-atom molecular dynamics. Int J Mol Sci, 2013. 14(5): p. 9893-905.
\bibitem{} Kmiecik, S., M. Jamroz, and A. Kolinski, Multiscale Approach to Protein Folding Dynamics, in Multiscale Approaches to Protein Modeling: Structure Prediction, Dynamics, Thermodynamics and Macromolecular Assemblies, A. Kolinski, Editor. 2011, Springer New York: New York, NY. p. 281-293.
\bibitem{} Horwacik, I., et al., Analysis and optimization of interactions between peptides mimicking the GD2 ganglioside and the monoclonal antibody 14G2a. Int J Mol Med, 2011. 28(1): p. 47-57.
\bibitem{} Kurcinski, M. and A. Kolinski, Hierarchical modeling of protein interactions. J Mol Model, 2007. 13(6-7): p. 691-8.
\bibitem{} Kurcinski, M. and A. Kolinski, Steps towards flexible docking: modeling of three-dimensional structures of the nuclear receptors bound with peptide ligands mimicking co-activators' sequences. J Steroid Biochem Mol Biol, 2007. 103(3-5): p. 357-60.
\bibitem{} Kurcinski, M. and A. Kolinski, Theoretical study of molecular mechanism of binding TRAP220 coactivator to Retinoid X Receptor alpha, activated by 9-cis retinoic acid. J Steroid Biochem Mol Biol, 2010. 121(1-2): p. 124-9.
\bibitem{} Kurcinski, M., A. Kolinski, and S. Kmiecik, Mechanism of Folding and Binding of an Intrinsically Disordered Protein As Revealed by ab Initio Simulations. J Chem Theory Comput, 2014. 10(6): p. 2224-31.
\bibitem{} Steczkiewicz, K., et al., Human telomerase model shows the role of the TEN domain in advancing the double helix for the next polymerization step. Proc Natl Acad Sci U S A, 2011. 108(23): p. 9443-8.
\bibitem{} Wabik, J., M. Kurcinski, and A. Kolinski, Coarse-Grained Modeling of Peptide Docking Associated with Large Conformation Transitions of the Binding Protein: Troponin I Fragment-Troponin C System. Molecules, 2015. 20(6): p. 10763-80.
\bibitem{} Kurcinski, M., M. Jamroz, and A. Kolinski, Multiscale Protein and Peptide Docking, in Multiscale Approaches to Protein Modeling: Structure Prediction, Dynamics, Thermodynamics and Macromolecular Assemblies, A. Kolinski, Editor. 2011, Springer New York: New York, NY. p. 21-33.
\bibitem{} Kurcinski, M., et al., CABS-dock web server for the flexible docking of peptides to proteins without prior knowledge of the binding site. Nucleic Acids Res, 2015. 43(W1): p. W419-24.
\bibitem{} Blaszczyk, M., et al., Modeling of protein-peptide interactions using the CABS-dock web server for binding site search and flexible docking. Methods, 2016. 93: p. 72-83.
\bibitem{} Webb, B. and A. Sali, Protein structure modeling with MODELLER. Methods Mol Biol, 2014. 1137: p. 1-15.
\bibitem{} Rajamani, D., et al., Anchor residues in protein-protein interactions. Proc Natl Acad Sci U S A, 2004. 101(31): p. 11287-92.
\bibitem{} Kabsch, W. and C. Sander, Dictionary of protein secondary structure: pattern recognition of hydrogen-bonded and geometrical features. Biopolymers, 1983. 22(12): p. 2577-637.
\end{thebibliography}
\end{document}